\pdfoutput=1

\documentclass[11pt,twoside,a4paper,cmspaper,final,collab]{cms-tdr}
\def\svnVersion{f0f8c59-D}\def\svnDate{2019/12/28}\def\cmsCernNoTag{CERN-EP-2019-135}\def\cmsCernDate{\today}\def\cmsMessage{"Published in Physical Review Letters as \href{http://dx.doi.org/10.1103/PhysRevLett.123.241801}{\doi{10.1103/PhysRevLett.123.241801}.}"}

\begin{document}\cmsNoteHeader{B2G-18-007}

\hyphenation{had-ron-i-za-tion}
\hyphenation{cal-or-i-me-ter}
\hyphenation{de-vices}
\RCS$Revision: 492000 $
\RCS$HeadURL: svn+ssh://svn.cern.ch/reps/tdr2/notes/B2G-18-007/trunk/B2G-18-007.tex $
\RCS$Id: B2G-18-007.tex 492000 2019-03-18 00:34:46Z skyriaco $

\newlength\cmsFigWidth
\ifthenelse{\boolean{cms@external}}{\setlength\cmsFigWidth{0.85\columnwidth}}{\setlength\cmsFigWidth{0.4\textwidth}}
\ifthenelse{\boolean{cms@external}}{\providecommand{\cmsLeft}{top\xspace}}{\providecommand{\cmsLeft}{left\xspace}}
\ifthenelse{\boolean{cms@external}}{\providecommand{\cmsRight}{bottom\xspace}}{\providecommand{\cmsRight}{right\xspace}}

\newcommand{\gammafr}{\ensuremath{f_{\gamma}}\xspace}

\title{Search for Physics beyond the Standard Model in Events with Overlapping Photons and Jets}

\date{\today}

\abstract{
Results are reported from a search for new particles that decay into a photon and two gluons, in events with jets. Novel jet substructure techniques are developed that allow photons to be identified in an environment densely populated with hadrons. The analyzed proton-proton collision data were collected by the CMS experiment at the LHC, in 2016 at ${\sqrt{s} = 13\TeV}$, and correspond to an integrated luminosity of 35.9\fbinv. The spectra of total transverse hadronic energy of candidate events are examined for deviations from the standard model predictions. No statistically significant excess is observed over the expected background. The first cross section limits on new physics processes resulting in such events are set. The results are interpreted as upper limits on the rate of gluino pair production, utilizing a simplified stealth supersymmetry model. The excluded gluino masses extend up to 1.7\TeV, for a neutralino mass of 200\GeV and exceed previous mass constraints set by analyses targeting events with isolated photons. 
}

\hypersetup{
  pdfauthor={CMS Collaboration},%
  pdftitle={Search for new physics in events with closely collimated photons and gluons},%
  pdfsubject={CMS},
  pdfkeywords={CMS, physics, analysis}}

\maketitle

Despite the success of the standard model (SM) of particle physics, there are a number of indications, such as the cosmological observations of dark matter and the low measured value of the Higgs boson mass, that suggest the existence of new physics at the TeV energy scale. No evidence for new physics has been uncovered thus far by the LHC. Signs of new phenomena could be hidden by high 
rate background SM processes that have yet to be properly explored. A large number of well-motivated theoretical scenarios predict the appearance of new physics in proton-proton collision events with low missing transverse momentum (\ptmiss) and nonisolated photons and leptons, which would appear as multijet events in a collider detector. These scenarios arise in  hidden valley models \cite{HiddenValleyEchoes,HiddenV} and a number of supersymmetric (SUSY) models, such as $R$ parity violating SUSY \cite{BARBIER20051} and stealth SUSY \cite{Fan2012,StealthSUSYnew,StealthSimple}. 

\begin{figure}[htbp]
  \centering
  \includegraphics[width=0.40\textwidth]{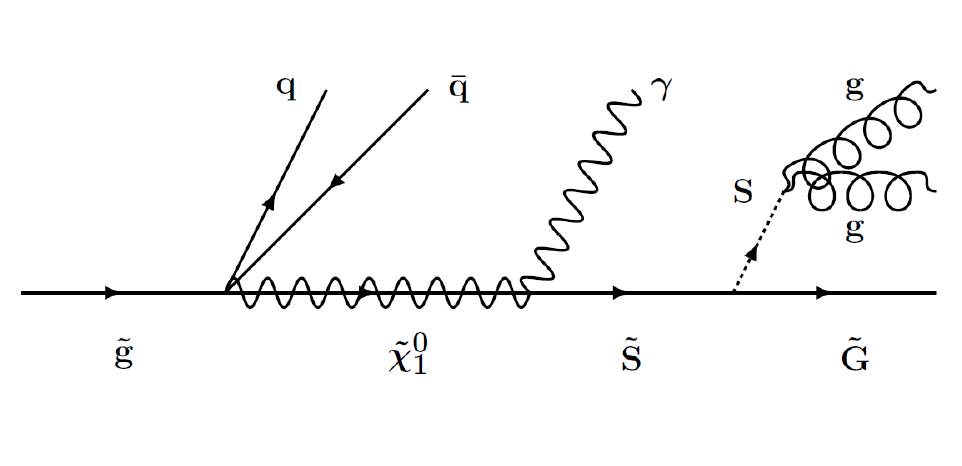}                                
    \caption{ The decay diagram for a single gluino as predicted by stealth SUSY. This analysis searches for pair produced gluinos and thus two such decay chains are expected in each signal event.}
  \label{fig:Diagram}

\end{figure}

Stealth SUSY predicts a hidden sector of particles with minimal couplings to the SUSY breaking mechanism. As a result, the superpartners in this sector are nearly mass degenerate. In the present analysis, a simplified stealth SUSY model is used as a benchmark. The model has only one light hidden sector superparticle pair, the singlino, and the singlet ($\widetilde{S}$ and $S$, respectively). Gluinos (\sGlu), the gluon superpartners, are expected to be created with large cross sections at the LHC and to decay to neutralinos \PSGczDo and a quark-antiquark pair. Stealth SUSY assumes gauginos (either neutralinos or charginos), which decay to a $\widetilde{S}$ and a photon (\PGg), to be the portal to the hidden-sector. The $\widetilde{S}$ is expected to decay to an $S$ and a massless gravitino (\sGra), with the subsequent decay of the $S$ to a pair of gluons. Because of the mass degeneracy of the hidden-sector pair, the \sGra is expected to be produced with low momentum and the event to be characterized by low \ptmiss. A diagram depicting the decay chain of a gluino according to this simplified stealth SUSY model is presented in Fig. \ref{fig:Diagram}. 

Previous searches at CMS for stealth SUSY \cite{CMS:Stealth,CMSStealthOld} required two isolated photons. The isolation requirement reduces the sensitivity for scenarios where a large mass difference exists between the electroweak gauginos, in this case the \PSGczDo and the colored superparticle (\sGlu). If this large mass interval is present, the \PSGczDo is expected to be produced with a large Lorentz boost and its decay products to be collimated, resulting in photons that are not isolated in the event. Since we search for events with jets composed of one photon from the \PSGczDo decay and a pair of gluons from the $S$ decay, which we refer to as \textit{photon jets}, our search is complementary to previous searches. It is possible to identify photon jets by utilizing a combination of existing and novel jet substructure tools. Within the simplified stealth SUSY model we consider, superparticles would be produced at the LHC in events with two photon jets associated with a large number of hadrons. The distribution of the total transverse hadronic energy of events containing photon jets is used to discriminate possible new physics obscured by the SM multijet background. 

The central feature of the CMS apparatus is a superconducting solenoid of 6\unit{m} internal diameter, providing a magnetic field of 3.8\unit{T}. Within the solenoid volume are a silicon pixel and strip tracker, a lead tungstate 
crystal electromagnetic calorimeter (ECAL), and a brass and scintillator hadron calorimeter, each composed of a barrel and two end cap sections. Forward calorimeters extend the pseudorapidity ($\eta$) coverage provided by the barrel and end cap detectors. Muons are detected in gas-ionization chambers embedded in the steel flux-return yoke outside the 
solenoid. Observed events that are considered potentially interesting are selected by a two-tiered trigger system \cite{hlt}. A more detailed description of the CMS detector, together with a definition of the coordinate system used and the relevant kinematic variables, can be found in Ref.~\cite{Chatrchyan:2008zzk}. 

Particle objects are reconstructed by the particle-flow algorithm \cite{PF}, from combinations of observations from the CMS detector components. The particle objects are clustered into jets using the anti-\kt algorithm \cite{antiKT} implemented in \FASTJET~\cite{fastjet} with a distance parameter of 0.8
(AK8 jets) and 0.4 (AK4 jets). The AK4 jet collection is utilized mainly for triggering purposes, while the larger radius AK8 jet collection, for the reconstruction of the \PSGczDo decays. The primary vertex is defined as the reconstructed vertex with the largest quadratic sum of the transverse momenta (\pt) of AK4 jets clustered from tracks associated with the vertex and the negative vector-\pt sum of these jets. Charged-particle candidates not associated
with the primary vertex are ignored to reduce pileup effects in the event reconstruction. Pileup refers to additional proton-proton (\Pp\Pp) collisions within the same or neighboring bunch crossings of the LHC beams. Jets are required to pass loose identification criteria \cite{jetid}, to reduce misreconstructed jets and jets reconstructed from calorimeter noise \cite{HCalnoise}. In addition, energy corrections are applied to the jets \cite{JEC}. Kinematic requirements of a minimum jet \pt of 200\GeV and the jet pseudorapidity $(\eta)$, to be $-2<\eta<2$ are applied to AK8 jets. The AK8 jet \pt is used to measure the total transverse hadronic activity in the event, defined as $\HT = \sum \pt$, where the sum is over all the AK8 jets in the event. For the analysis, we consider events that have $\HT>1\TeV$ and contain at least 3 AK8 jets. 

The data analyzed were collected by the CMS experiment at the LHC from \Pp\Pp{} collisions at ${ \sqrt{s} = 13\TeV }$ during the 2016 data taking period, and correspond to an integrated luminosity of 35.9\fbinv. Events are selected by the trigger system if they pass a minimum \HT requirement of 900\GeV, calculated using the AK4 jets with a minimum \pt of 50\GeV and $\abs{\eta}<2.5$. For the purpose of correcting data-to-simulation differences, events were also collected with a combination of muon triggers, selecting events containing at least one muon with \pt greater than 50\GeV.

Pair production of gluinos for a range of different \sGlu and \PSGczDo masses, with the $S$ and $\widetilde{S}$ masses fixed to 90 and 100\GeV, respectively, are simulated using \MGvATNLO \cite{MADGRAPH}. The decay and hadronization are done with \PYTHIA \cite{PYTHIA} using the \textsc{CUETP8M1} tune \cite{CTune} for the underlying event and the \textsc{NNPDF3.0} parton distribution functions (PDF) \cite{nnpdf}. The detector is simulated with the CMS fast simulation package (\textsc{FastSIM}) \cite{CMSFastSIM,FASTSIM2}. To estimate systematic uncertainties related to the detector simulation, the full CMS detector simulation (\textsc{FullSIM}) based on \GEANTfour \cite{GEANT4} is also used and its results are compared to those of \textsc{FastSIM}. An uncertainty due to the hadronization model is evaluated by an alternative signal simulation with \HERWIG \cite{herwig} and the \textsc{TUNEEE5C} \cite{herwig2} underlying event tune. Signal events are normalized using the theoretical gluino pair production cross sections \cite{CrossSectionsGluinos} at next-to-leading order, assuming a 100\% branching fraction to the \sGlu decay channel shown in Fig. \ref{fig:Diagram}. 

We simulate SM processes to study the behavior of the background, to construct templates from which we estimate the efficiency corrections used for simulated signals, and to estimate the various uncertainties. The dominant background is from quantum chromodynamic (QCD) multijet processes. Simulation of QCD processes is done using \MGvATNLO with MLM matching \cite{madMLM} and hadronized with \PYTHIA8 with the \textsc{CUETP8M1} tune. The production of hadronically and leptonically decaying \PW bosons in conjunction with jets (\PW{}+jets) is also simulated this way. Top quark-antiquark pairs (\ttbar) are simulated with \POWHEG{}v2 \cite{powheg1,powheg2,powheg3,powheg4} and hadronized by \PYTHIA{8} using the \textsc{CUETP8M2T4} \cite{CTune} underlying event tune. As an alternative to \PYTHIA, \HERWIG with the \textsc{TUNEEE5C} underlying event tune are also used for hadronization of \ttbar pairs. All samples are simulated with the \textsc{NNPDF3.0} PDFs. The detector response is simulated using \GEANTfour.     

\begin{figure}[htbp]
  \centering
  \includegraphics[width=0.50\textwidth]{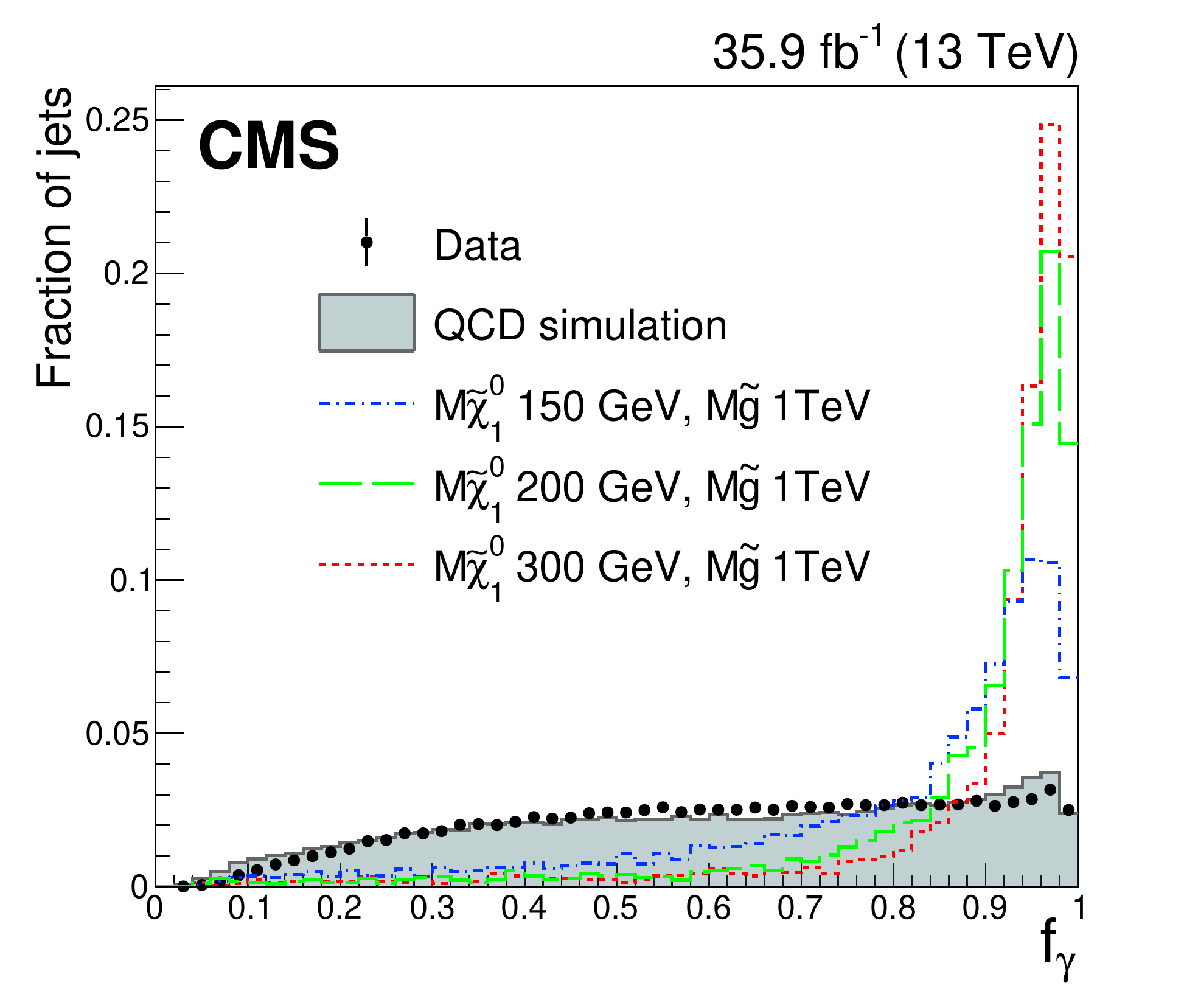}
    \caption{Distribution of the photon subjet energy fraction (\gammafr) for jets that satisfy the loose photon jet requirements. Simulated distributions for signal are denoted by the broken lines, each depicting a different mass of \PSGczDo and \sGlu. The shaded area represents the QCD jets distribution. }
  \label{fig:subjf}
\end{figure}

Each AK8 jet in the event is examined to identify candidate photon jets, which will have a three-prong substructure and a photon from the \PSGczDo decay. We require that there is at least one photon cluster in the AK8 jet, with $\pt>20\GeV$ and at least $95\%$ of the energy deposited in ECAL, consistent with a photon shower shape \cite{Photons8tev}. This photon candidate is also required to not have any associated hits in the pixel detector (pixel veto). Photons converting in the tracker material can produce multiple PF objects, which are replaced by the reconstructed photon object four vector. The photon and the AK8 jet constituents are reclustered using the \kt algorithm \cite{KTalgo} and the merging history is examined to identify the three subjets of the jet. The clustering algorithm combines two objects into one at each step. We identify as the first subjet, the less massive of the two objects merged in the last step of the clustering sequence. The other object, the more massive of the two, specifies the $2^{nd}$ and $3^{rd}$ subjets. To be considered a photon jet, the AK8 jet must have three subjets with $\pt>10\GeV$. We further examine the subjet that contains the photon and define the photon subjet energy fraction (\gammafr)~ as the ratio of the photon's transverse energy to the subjet's \pt. The \gammafr distribution is shown in Fig. \ref{fig:subjf} for data, simulated multijet backgrounds, and simulated signal. This variable is a measure of the activity around the photon and serves as a strong discriminator against the QCD multijet background. 

An additional jet-substructure tool is used to enhance the discrimination between signal like three-prong jets, and background dominated single prong jets. In this approach the \textit{N}-subjettiness variables \cite{nsubjettiness} denoted by $\tau_{N}$ are used to determine the consistency of a jet with \textit{N} or fewer prongs. The $\tau_{N}$ values are defined as the following:
                                \begin{linenomath}
                                \begin{equation}
                                \begin{aligned}
                                 \tau_{N}  \equiv \frac{1}{d_{0}} \sum\limits_{i}{\pt}_{i}min \{ \Delta R_{1,i},\Delta R_{2,i} ....\Delta R_{N,i} \} 
                                \end{aligned}
                                \end{equation}
                                \end{linenomath}
where the index $i$ refers to each jet constituent, $\Delta R$ is the angular distance between a jet constituent and a candidate subjet axis, and $d_{0}$ is a normalization constant.  Jets composed of three subjets should have small values for the ratio $\tau_{3}/\tau_{1}$. Photon jets are required to satisfy the condition $\tau_{3}/\tau_{1}<0.4$. Photon jets satisfying the additional requirement $\gammafr>0.9$ are categorized as \textit{tight} photon jets, the rest are referred to as \textit{loose} photon jets. Events are characterized by their multiplicity of loose and tight photon jets, and are labeled as \textit{X-Y} where \textit{X} is the number of loose photon jets, of which \textit{Y} also satisfies the tight photon jet criteria. 
We define the signal region (SR) as that containing events with exactly two loose photon jets, while the background dominated region (BR) contains events with one or less loose photon jet. The SR is further split into three multiplicity categories, 2-0, 2-1, and 2-2, with the last one being the most sensitive to the signal.

The SM multijet background is estimated from data. The probabilities for a QCD jet to be labeled as a loose or tight photon jet, referred to as mistag rates, are measured in the BR as a function of the jet \pt and $\eta$. The loose mistag rate is measured by taking the ratio of the number of jets passing the loose selection in the BR, to the total number of the jets in the BR, as a function of 
jet \pt and $\eta$. The tight photon jet mistag rate is the ratio of the number of tight photon jets to the number of all loose photon jets in the BR. The probabilities of each event to populate the three SR categories are calculated by generating an ensemble of $10^4$ pseudoexperiments for each event in the BR, using the AK8 jet kinematic variables and the measured mistag rates. One can then obtain the background \HT distributions, for each SR category. This is achieved by constructing an \HT distribution of all events in the BR and weighting each event by the calculated probabilities for it to pass the SR selections. The mistag rates are varied within their statistical uncertainties to determine the uncertainty in the background prediction. It was found that the background contribution is underestimated in events where overlap between neighboring jets exists. Therefore, in each event, the minimum pairwise distance in the $\eta\text{--}\phi$ space between AK8 jets, defined as $\Delta R=\sqrt{\smash[b]{(\Delta \eta)^2 + (\Delta \phi)^2}}$, is required to be $\Delta R{>}1.5$. The validity of the background method is tested by confirming that there is agreement between prediction and observation for the numbers of events in the \HT distributions and for their shapes. The tests are performed both with simulated events and with a subset of the data corresponding to 10\% of the total integrated luminosity. In each case the method is found to achieve closure to within 5\%. Other SM processes such as \ttbar and \PW{}+jets are simulated and estimated to have a negligible contribution in the SR.  

To measure the signal efficiency correction for the loose and tight photon jet selections, since no SM process predicts jets composed of a collimated photon and two gluons, we select AK8 jets that are composed of an electron, a bottom quark and a final-state radiation gluon, originating from top quark decays. This approach requires the pixel veto constraint to be reversed in order to allow an electron in a jet to emulate a photon. A \ttbar dominated sample is selected by tagging events in which the combination of a muon, a loosely \PQb-tagged  AK4 jet \cite{btag} and \ptmiss is back to back to an AK8 jet (probe jet). The probe jets are used for the measurement of the loose and tight photon jet rates. The measurement is done by fitting simulation-based templates to the probe jets, estimating the data composition (e.g., jets originating from light quarks or gluons, or fully merged hadronic \PW~boson or top quark decays) and measuring the loose and tight photon jets selection efficiency. The procedure is repeated in simulation and the efficiency correction is defined as the ratio of the loose or tight efficiency measured in data over the one obtained from \ttbar simulation. The templates are constructed using the probe jet mass. Using simulated top pairs and signal samples hadronized with \PYTHIA and \HERWIG, an uncertainty is derived to address the differences in the jet constituents between top and signal jets. Finally, the signal yield is scaled to correct for the difference between data and simulation, and the associated uncertainty in the yield is estimated by measuring the impact of changing the scaling factor by its uncertainty. 

The dominant source of systematic uncertainty is the data-to-simulation efficiency correction for signal-like jets. This ranges from 30 to 50$\%$ depending on the event jet composition. The uncertainties considered and their magnitudes are listed in Table \ref{tab:SystematicUnc}. These include uncertainties associated with the following sources: background estimation, jet calibration and resolution corrections, which can affect the measured jet energy \cite{cms:jets}, pileup modeling, the total integrated luminosity measurement \cite{CMS:lumi}, simulation effects for signal such as the difference between the full and fast detector simulation, and the PDF choice \cite{PDF4LHCrec}. Initial-state radiation effects on signal efficiency and triggering efficiency uncertainties are estimated to be negligible and not included. Systematic uncertainties are introduced as shape or normalization variations for the limit setting procedure, as indicated in Table \ref{tab:SystematicUnc}.

\begin{table}
 \centering
 \topcaption{ Impact of systematic uncertainties on either signal acceptance ($^s$) or background($^b$) . Shape uncertainties are denoted by an asterisk ($^\ast$), while the others are considered normalization uncertainties.}  
 \begin{tabular}{ l   r  }
     \hline
     Source  & Impact \\
     \hline 
     Simulation-to-data signal efficiency correction $^{\ast s}$ & $30\text{--}50\%$  \\

     Background estimation $^{\ast b}$ &$5\text{--}20\%$ \\

     Jet energy resolution $^{\ast sb}$ &${<}10\%$ \\
     
     Jet energy scale corrections $^{\ast sb}$ &${<}10\%$ \\
         
     Pileup reweighting $^{\ast s}$ &${<}5\%$     \\  
     
     Integrated luminosity$^{s}$ &$2.5\%$  \\
     
     Detector \textsc{FullSIM--FastSIM}$^{s}$ &$1\text{--}2\%$ \\

     PDF choice uncertainty$^{s}$ &$1\%$ \\
     \hline

 \end{tabular}
\label{tab:SystematicUnc}
  
\end{table}

\begin{figure*}[htbp]
  \centering
  \includegraphics[width=0.95\textwidth]{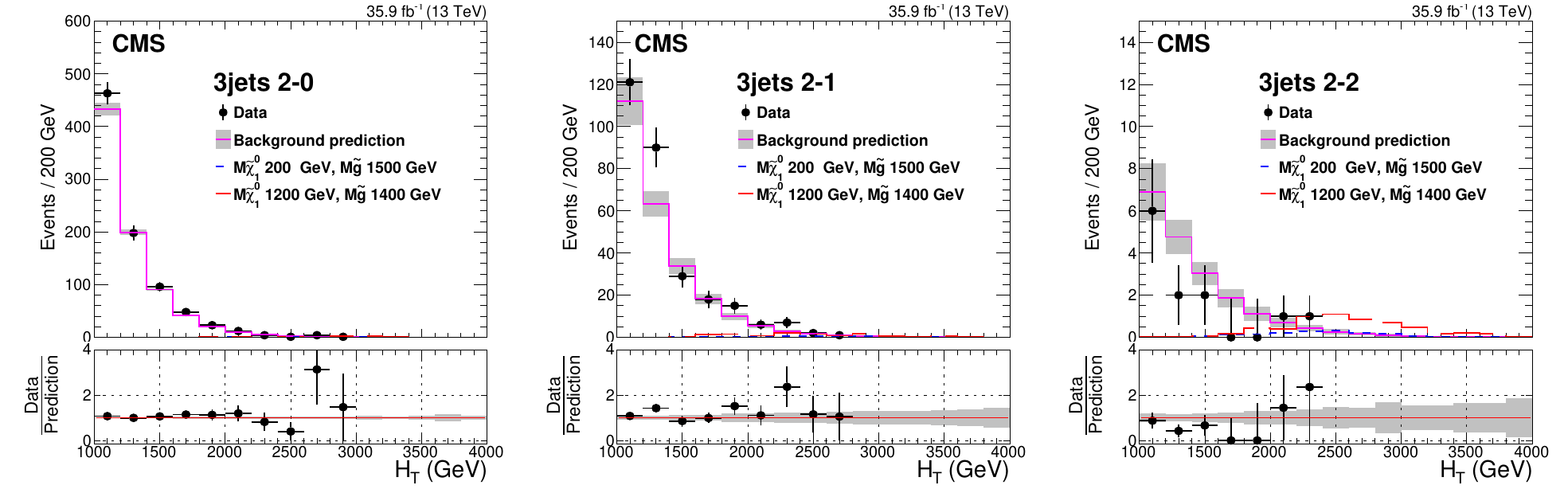}\\
  \includegraphics[width=0.95\textwidth]{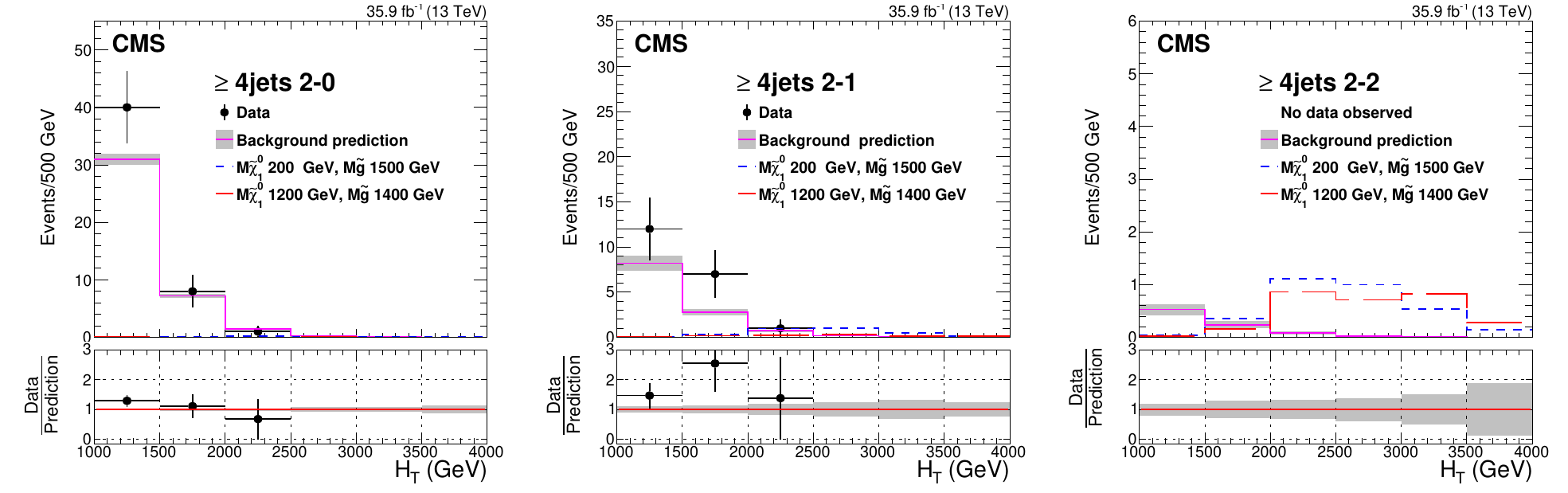}
  \caption{The \HT distributions in the signal regions for the 3-jet AK8 (upper row) and the $\geq$ 4 AK8 jets categories (lower row). Events with zero, one and two tight photon jets are presented from left to right. The magenta line with the gray band corresponds to the background expectation obtained from data while the blue and red colored lines present two signal benchmarks. The lower panels present the data-to-background ratio with their respective uncertainties.} 
  \label{fig:HTplots}
\end{figure*}

\begin{figure}[htbp]
  \centering
  \includegraphics[width=0.50\textwidth]{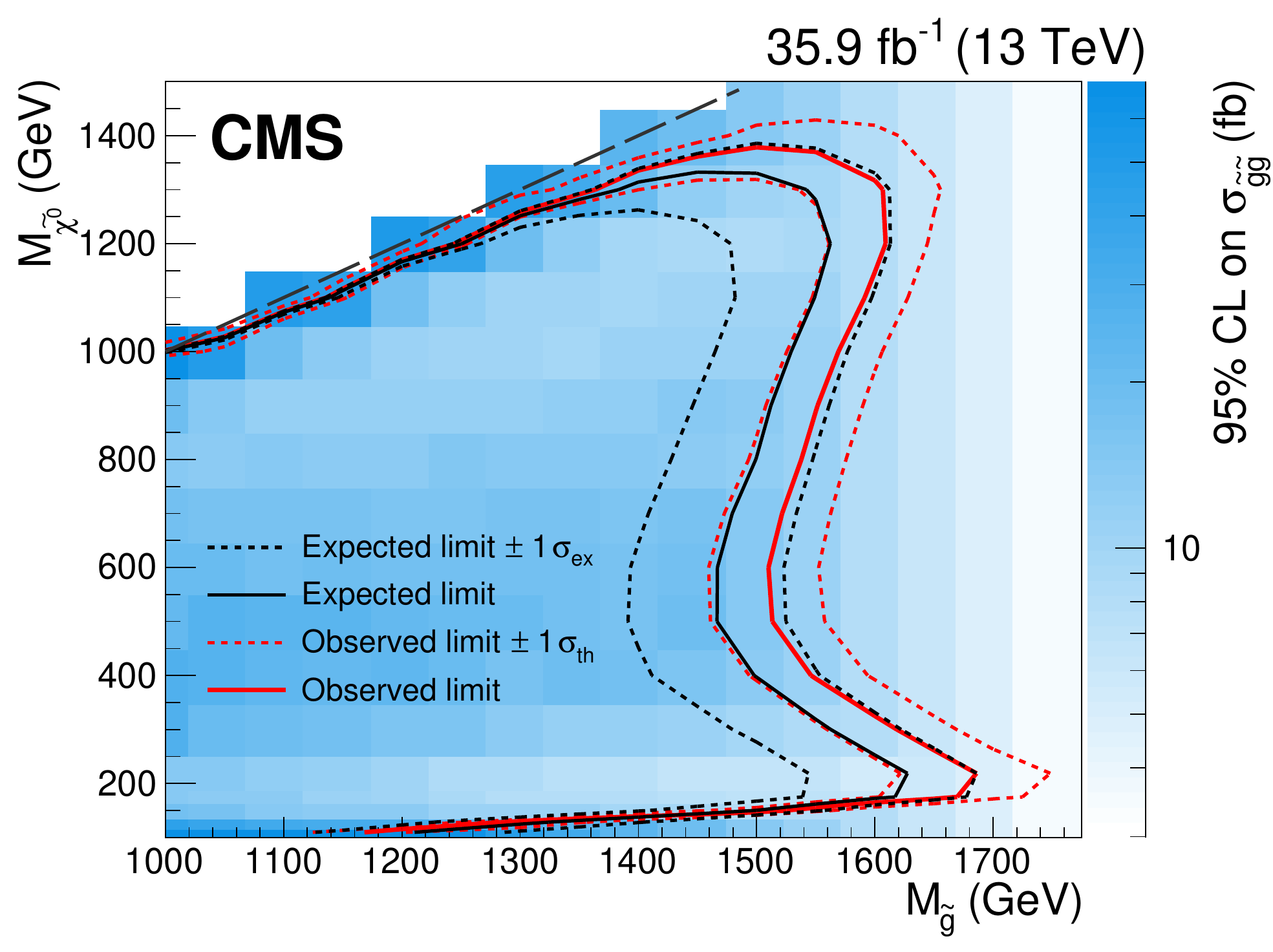}
    \caption{ The upper limit at 95\% confidence level on the \sGlu pair production cross section as a function of \sGlu and \PSGczDo masses. The region enclosed by the red (lighter) solid line is excluded. The black (darker) solid line presents the expected excluded area. The uncertainty in the observed limit corresponds to the theoretical uncertainties in the signal cross section. Exclusion in the low \PSGczDo and high \sGlu mass region is a result of the implementation of the substructure techniques.}
  \label{fig:Cov}
\end{figure}

The search is performed separately on events with exactly three AK8 jets and events with four or more AK8 jets. A joint statistical analysis is performed using the \HT spectra in the six SR considered. The \HT distributions in the SR are presented in Fig. \ref{fig:HTplots}, where it can be seen that the data are consistent with the background prediction. We interpret the results as upper limits on the cross section for pair-produced gluinos, decaying according to the simplified stealth SUSY model, using a Bayesian limit setting method with a flat signal prior \cite{CowanPDGStat}. The systematic uncertainties are incorporated as nuisance parameters with log-normal priors and are assumed to be correlated among the six SR. The cross section limits for all SR categories are shown in Fig. \ref{fig:Cov}. Production of \sGlu with masses up to 1.7\TeV are excluded at a 95\% confidence level, for an assumed \PSGczDo mass of 200\GeV. For neutralino masses between 1.0 and 1.2\TeV, the maximum excluded gluino mass is 1.5\text{--}1.7\TeV. This is the first result on boosted final states with photons and gluons merging into a single jet. The resulting limits improve over those obtained in previous analyses searching for isolated photons.

To summarize, a search for new particles decaying to a photon and two gluons in events with jets is presented. The search is performed in events with two jets that have substructure and are composed of a photon and two gluons. A dataset of proton-proton collisions at a center-of-mass energy of 13\TeV collected by the CMS experiment, corresponding to an integrated luminosity of 35.9\fbinv, is analyzed. To identify the candidate jets, novel jet substructure techniques have been developed and used to complement established methods. The total transverse hadronic activity distributions of events in the signal region are compared to the expected distributions, estimated from data. No statistically significant excess is observed above the standard model background expectation. We establish  upper limits at 95\% confidence level on the cross section for gluino pair production, using a simplified stealth SUSY model. The excluded gluino masses extend up to 1.5--1.7\TeV, depending on the neutralino mass, with the highest exclusion set for neutralinos with a mass of 200\GeV. This is the first search of this kind targeting the region of parameter space where photons from neutralino decays are not isolated.

\begin{acknowledgments}
We congratulate our colleagues in the CERN accelerator departments for the excellent performance of the LHC and thank the technical and administrative staffs at CERN and at other CMS institutes for their contributions to the success of the CMS effort. In addition, we gratefully acknowledge the computing centers and personnel of the Worldwide LHC Computing Grid for delivering so effectively the computing infrastructure essential to our analyses. Finally, we acknowledge the enduring support for the construction and operation of the LHC and the CMS detector provided by the following funding agencies: BMBWF and FWF (Austria); FNRS and FWO (Belgium); CNPq, CAPES, FAPERJ, FAPERGS, and FAPESP (Brazil); MES (Bulgaria); CERN; CAS, MoST, and NSFC (China); COLCIENCIAS (Colombia); MSES and CSF (Croatia); RPF (Cyprus); SENESCYT (Ecuador); MoER, ERC IUT, PUT and ERDF (Estonia); Academy of Finland, MEC, and HIP (Finland); CEA and CNRS/IN2P3 (France); BMBF, DFG, and HGF (Germany); GSRT (Greece); NKFIA (Hungary); DAE and DST (India); IPM (Iran); SFI (Ireland); INFN (Italy); MSIP and NRF (Republic of Korea); MES (Latvia); LAS (Lithuania); MOE and UM (Malaysia); BUAP, CINVESTAV, CONACYT, LNS, SEP, and UASLP-FAI (Mexico); MOS (Montenegro); MBIE (New Zealand); PAEC (Pakistan); MSHE and NSC (Poland); FCT (Portugal); JINR (Dubna); MON, RosAtom, RAS, RFBR, and NRC KI (Russia); MESTD (Serbia); SEIDI, CPAN, PCTI, and FEDER (Spain); MOSTR (Sri Lanka); Swiss Funding Agencies (Switzerland); MST (Taipei); ThEPCenter, IPST, STAR, and NSTDA (Thailand); TUBITAK and TAEK (Turkey); NASU and SFFR (Ukraine); STFC (United Kingdom); DOE and NSF (USA).
\end{acknowledgments}

\bibliography{auto_generated}

\cleardoublepage \appendix\section{The CMS Collaboration \label{app:collab}}\begin{sloppypar}\hyphenpenalty=5000\widowpenalty=500\clubpenalty=5000\vskip\cmsinstskip
\textbf{Yerevan Physics Institute, Yerevan, Armenia}\\*[0pt]
A.M.~Sirunyan$^{\textrm{\dag}}$, A.~Tumasyan
\vskip\cmsinstskip
\textbf{Institut f\"{u}r Hochenergiephysik, Wien, Austria}\\*[0pt]
W.~Adam, F.~Ambrogi, T.~Bergauer, J.~Brandstetter, M.~Dragicevic, J.~Er\"{o}, A.~Escalante~Del~Valle, M.~Flechl, R.~Fr\"{u}hwirth\cmsAuthorMark{1}, M.~Jeitler\cmsAuthorMark{1}, N.~Krammer, I.~Kr\"{a}tschmer, D.~Liko, T.~Madlener, I.~Mikulec, N.~Rad, J.~Schieck\cmsAuthorMark{1}, R.~Sch\"{o}fbeck, M.~Spanring, D.~Spitzbart, W.~Waltenberger, C.-E.~Wulz\cmsAuthorMark{1}, M.~Zarucki
% ...
\vskip\cmsinstskip
\dag: Deceased\\
1:  Also at Vienna University of Technology, Vienna, Austria\\
% ...
\end{sloppypar}
\end{document}